\RequirePackage{lineno}

\documentclass[aps,prc,twocolumn,groupedaddress,showpacs,amsmath,amssymb,floatfix,superscriptaddress]{revtex4}
\usepackage{multirow}
\usepackage{graphicx}
\usepackage{times}
\usepackage{color}

\begin{document}

\bibliographystyle{h-physrev3}

\title{Reactor On-Off Antineutrino Measurement with KamLAND}

\newcommand{\tohoku}{\affiliation{Research Center for Neutrino
    Science, Tohoku University, Sendai 980-8578, Japan}}
\newcommand{\osaka}{\affiliation{Graduate School of 
    Science, Osaka University, Toyonaka, Osaka 560-0043, Japan}}
\newcommand{\alabama}{\affiliation{Department of Physics and
    Astronomy, University of Alabama, Tuscaloosa, Alabama 35487, USA}}
\newcommand{\lbl}{\affiliation{Physics Department, University of
    California, Berkeley, and \\ Lawrence Berkeley National Laboratory, 
Berkeley, California 94720, USA}}
\newcommand{\colostate}{\affiliation{Department of Physics, Colorado
    State University, Fort Collins, Colorado 80523, USA}}
\newcommand{\hawaii}{\affiliation{Department of Physics and Astronomy,
    University of Hawaii at Manoa, Honolulu, Hawaii 96822, USA}}
\newcommand{\ut}{\affiliation{Department of Physics and
    Astronomy, University of Tennessee, Knoxville, Tennessee 37996, USA}}
\newcommand{\tunl}{\affiliation{Triangle Universities Nuclear Laboratory, Durham, North Carolina 27708, USA; Physics Departments at Duke University, Durham, North Carolina 27705, USA; North Carolina Central University, Durham, North Carolina 27701, USA and the University of North Carolina at Chapel Hill, Chapel Hill, North Carolina 27599, USA}}
\newcommand{\ipmu}{\affiliation{Kavli Institute for the Physics and Mathematics of the Universe (WPI),
University of Tokyo, Kashiwa, 277-8583, Japan}}
\newcommand{\nikhef}{\affiliation{Nikhef and the University of Amsterdam, Science Park, Amsterdam, the Netherlands}}
\newcommand{\washington}{\affiliation{Center for Experimental Nuclear Physics and Astrophysics, University of Washington, Seattle, Washington 98195, USA}}

\newcommand{\OscPhaseEff}{\Delta m^{2}\mathbb{L}/\mathbb{E})_{\rm{Eff}}}

%
%
\author{A.~Gando}\tohoku
\author{Y.~Gando}\tohoku
\author{H.~Hanakago}\tohoku
\author{H.~Ikeda}\tohoku
\author{K.~Inoue}\tohoku\ipmu
\author{K.~Ishidoshiro}\tohoku
\author{H.~Ishikawa}\tohoku
\author{M.~Koga}\tohoku\ipmu
\author{R.~Matsuda}\tohoku
\author{S.~Matsuda}\tohoku
\author{T.~Mitsui}\tohoku
\author{D.~Motoki}\tohoku
\author{K.~Nakamura}\tohoku\ipmu
\author{A.~Obata}\tohoku
\author{A.~Oki}\tohoku
\author{Y.~Oki}\tohoku
\author{M.~Otani}\tohoku
\author{I.~Shimizu}\tohoku
\author{J.~Shirai}\tohoku
\author{A.~Suzuki}\tohoku
\author{Y.~Takemoto}\tohoku
\author{K.~Tamae}\tohoku
\author{K.~Ueshima}\tohoku
\author{H.~Watanabe}\tohoku
\author{B.D.~Xu}\tohoku
\author{S.~Yamada}\tohoku
\author{Y.~Yamauchi}\tohoku
\author{H.~Yoshida}\tohoku

\author{A.~Kozlov}\ipmu

\author{S.~Yoshida}\osaka

\author{A.~Piepke}\ipmu\alabama

\author{T.I.~Banks}\lbl
\author{B.K.~Fujikawa}\ipmu\lbl
\author{K.~Han}\lbl
\author{T.~O'Donnell}\lbl

\author{B.E.~Berger}\colostate

\author{J.G.~Learned}\hawaii
\author{S.~Matsuno}\hawaii
\author{M.~Sakai}\hawaii

\author{Y.~Efremenko}\ipmu\ut

\author{H.J.~Karwowski}\tunl
\author{D.M.~Markoff}\tunl
\author{W.~Tornow}\ipmu\tunl

\author{J.A.~Detwiler}\washington
\author{S.~Enomoto}\ipmu\washington

\author{M.P.~Decowski}\ipmu\nikhef

\collaboration{The KamLAND Collaboration}\noaffiliation

\date{\today}

\begin{abstract}
The recent long-term shutdown of Japanese nuclear reactors has resulted in a significantly reduced reactor $\overline{\nu}_{e}$ flux at KamLAND. This running condition provides a unique opportunity to confirm and constrain backgrounds for the reactor $\overline{\nu}_{e}$ oscillation analysis. The data set also has improved sensitivity for other $\overline{\nu}_{e}$ signals, in particular  $\overline{\nu}_{e}$'s produced in $\beta$-decays from $^{238}$U and $^{232}$Th within the Earth's interior, whose energy spectrum overlaps with that of reactor $\overline{\nu}_{e}$'s. Including constraints on $\theta_{13}$ from accelerator and short-baseline reactor neutrino experiments, a combined three-flavor analysis of solar and KamLAND data gives fit values for the oscillation parameters of $\tan^{2} \theta_{12} = 0.436^{+0.029}_{-0.025}$, $\Delta m^{2}_{21} = 7.53^{+0.18}_{-0.18} \times 10^{-5}\,{\rm eV}^{2}$, and \mbox{$\sin^{2} \theta_{13} = 0.023^{+0.002}_{-0.002}$}. Assuming a chondritic Th/U mass ratio, we obtain $116^{+28}_{-27}$ $\overline{\nu}_{e}$ events from $^{238}$U and $^{232}$Th, corresponding to a geo $\overline{\nu}_{e}$ flux of \mbox{$3.4^{+0.8}_{-0.8} \times 10^{6}\,{\rm cm^{-2}s^{-1}}$} at the KamLAND location. We evaluate various bulk silicate Earth composition models using the observed  geo $\overline{\nu}_{e}$ rate.
\end{abstract}


\maketitle

\section{Introduction}
\label{section:Introduction}

\vspace{-0.2cm}

The Kamioka Liquid-scintillator Antineutrino Detector (KamLAND) demonstrated the oscillatory nature of neutrino flavor transformation by observing electron antineutrinos ($\overline{\nu}_{e}$) with energies of a few MeV from nuclear reactors typically 180\,km away~\cite{Gando2011a}. Following the Fukushima nuclear accident in March 2011, the entire Japanese nuclear reactor industry, which generates $>$97\% of the reactor $\overline{\nu}_{e}$ flux at KamLAND, has been subjected to a protracted shutdown due to a review of nuclear safety standards. 
This unexpected situation allows for a reactor on-off study of backgrounds for the KamLAND neutrino oscillation analysis.  

The reactor-off data also yield improved sensitivity for $\overline{\nu}_{e}$'s produced by other sources. Previously, we used the KamLAND data to search for geoneutrinos, $\overline{\nu}_{e}$'s produced in $\beta$-decays from primordial radioactivity within the Earth's interior. The $^{238}$U and $^{232}$Th decay chains emit $\overline{\nu}_{e}$'s with energies below 3.4\,MeV, so reactor $\overline{\nu}_{e}$ events with similar energies pose a background for this signal.  Despite having a high reactor  $\overline{\nu}_{e}$ background, KamLAND performed  the first experimental study of geo $\overline{\nu}_{e}$'s from the decay chains of $^{238}$U and $^{232}$Th~\cite{Araki2005b}. Later the geo  $\overline{\nu}_{e}$ signal at KamLAND was used to estimate our planet's radiogenic heat production and constrain composition models of the bulk silicate Earth (BSE, the Earth's region outside its metallic core). In particular it was found that fully-radiogenic Earth models are disfavored ~\cite{Gando2011b}. The Borexino experiment at Gran Sasso also reported a positive observation of geo $\overline{\nu}_{e}$'s~\cite{Bellini2010}.

In this article, we present improved reactor neutrino oscillation results and geo $\overline{\nu}_{e}$ flux measurements including the recent reactor-off period. For the reactor $\overline{\nu}_{e}$ rate estimate, we also apply new evaluations of reactor antineutrino emission spectra, as well as constraints on oscillation parameters from accelerator and short-baseline reactor neutrino oscillation measurements.

\section{Neutrino oscillation}
\label{section:NeutrinoOscillation}

Neutrino oscillation is well established by experimental studies of solar, reactor, atmospheric, and accelerator neutrinos. KamLAND  observes $\overline{\nu}_{e}$'s from many reactors at a flux-weighted average distance of 180\,km, providing optimal sensitivity for the LMA-MSW ${\nu}_{1}$-${\nu}_{2}$ mixing solution to the solar neutrino problem. For the length scale relevant to reactor $\overline{\nu}_{e}$ oscillation at KamLAND, the three-flavor survival probability ($P^{3\nu}_{ee}$), including matter effects, may be approximated as 
\begin{eqnarray}
\label{equation:survival_probability_3nu}
P_{ee}^{3\nu} = \cos^{4}\theta_{13} \widetilde{P}_{ee}^{2\nu} + \sin^{4}\theta_{13}\;.
\end{eqnarray}
The two neutrino survival probability $\widetilde{P}_{ee}^{2\nu}$ has the same form as the survival probability in matter for \mbox{$\nu_{1}$-$\nu_{2}$} mixing but with the electron density \mbox{($N_{e}$)} modified: \mbox{$\widetilde{N}_{e} = N_{e}\cos^{2}\theta_{13}$}~\cite{Goswami2005}. It is given by
\begin{eqnarray}
\label{equation:survival_probability_2nu}
\widetilde{P}_{ee}^{2\nu} = 1 - \sin^{2} 2\theta_{12M} \sin^{2} \left( \frac{\Delta m_{21M}^{2} L}{4 E_{\nu}} \right), \label{Eq-P-2nu-KamLAND}
\end{eqnarray}
where $L$ is the distance from the source to the detector, $E_{\nu}$ is the $\overline{\nu}_{e}$ energy, and $\theta_{12M}$ and $\Delta m^2_{21M}$ are the matter-modified mixing angle and mass splitting defined by
\begin{eqnarray}
\label{Eq-Theta12M}
\sin^{2} 2\theta_{12M} = \frac{\sin^{2} 2\theta_{12}}{(\cos 2\theta_{12} - A / \Delta m_{21}^{2})^{2} + \sin^{2} 2\theta_{12}}\;,\\ 
\Delta m_{21M}^{2} = \Delta m_{21}^{2} \sqrt{(\cos 2\theta_{12} - A / \Delta m_{21}^{2})^{2} + \sin^{2} 2\theta_{12}}\;.
\end{eqnarray}
The parameter $A = \pm2 \sqrt{2} G_{F} \widetilde{N}_{e} E_{\nu}$ has a negative sign for antineutrinos; $G_{F}$ is the Fermi coupling constant.

Recently, accelerator and short-baseline ($\sim$1\,km) reactor experiments have demonstrated that $\theta_{13}$ is non-zero, and have measured it with high precision~\cite{Abe2011,Adamson2011,Abe2012,An2013,Ahn2012}. An analysis incorporating this new $\theta_{13}$ constraint will improve the determination of the other oscillation parameters.

\section{Geoneutrino flux at KamLAND}
\label{section:Geoneutrino}

While the mechanical properties of the Earth's interior are well established, its composition, including its radiochemical content, remains uncertain. Decays of uranium (U), thorium (Th), potassium (K) and their progeny generate heat.  Depending on their abundance and distribution within the Earth, these decays may be an essential heat source for generating Earth dynamics. A leading BSE model~\cite{McDonough1995} based on measured elemental abundances of chondritic meteorites and mantle peridotites predicts a radiogenic heat production of 8\,TW from the $^{238}$U decay chain, 8\,TW from the $^{232}$Th decay chain, and 4\,TW from $^{40}$K~\cite{Arevalo2009}. This would account for nearly half of the heat dissipation rate from the Earth's surface, which a recent analysis finds to be $47 \pm 2$\,TW~\cite{Davies2010}.

The energy spectrum of $^{40}$K neutrinos falls entirely below the 1.8~MeV energy threshold for the inverse $\beta$-decay reaction by which KamLAND observes antineutrinos, rendering these decays invisible to KamLAND.  However, the $^{238}$U and $^{232}$Th decay chain $\bar{\nu}_{e}$'s extend above this threshold with distinct energy distributions, making possible a direct measurement of the individual $^{238}$U and $^{232}$Th contributions.

The geo $\overline{\nu}_{e}$ flux at the KamLAND detector can be calculated from the isotope abundances $a_i(\vec{r'})$ for each isotope $i$ at source positions $\vec{r'}$ by integrating over the entire Earth, 
\begin{eqnarray}
\label{equation:flux}
\frac{\mathrm{d}\Phi (E_{\nu}, \vec{r})}{\mathrm{d}E_{\nu}} \hspace{6.0cm} \nonumber \\
 = \sum_{i} A_{i} \frac{\mathrm{d}n_{i} (E_{\nu})}{\mathrm{d}E_{\nu}} \int_{\oplus} \mathrm{d}^{3}\vec{r'} \frac{a_{i}(\vec{r'}) \rho(\vec{r'}) P_{ee}(E_{\nu}, |\vec{r} - \vec{r'}|)}{4\pi|\vec{r} - \vec{r'}|^{2}}, 
\end{eqnarray}
where $\vec{r}$ is the detector position, $A_{i}$ is the decay rate per unit mass, ${\mathrm{d}n_{i} (E_{\nu})}/{\mathrm{d}E_{\nu}}$ is the $\overline{\nu}_{e}$ energy spectrum for each mode of decay, $a_{i}(\vec{r'})$ is the isotope mass per unit rock mass, $\rho(\vec{r'})$ is the rock density, and $P_{ee}(E_{\nu}, |\vec{r} - \vec{r'}|)$ is the $\overline{\nu}_{e}$ survival probability given by Eq. (\ref{equation:survival_probability_3nu}) with $L = |\vec{r} - \vec{r'}|$. Given the measured values of neutrino oscillation parameters and the energy range of detectable geo $\overline{\nu}_{e}$'s, the integration over the volume of the Earth averages over the second sine function in Eq.~(\ref{equation:survival_probability_2nu}), allowing the approximation, 
\begin{eqnarray}
\label{equation:oscillation2}
P_{ee}^{3\nu} \simeq \cos^{4}\theta_{13} \left( 1 - \frac{1}{2} \sin^{2}2\theta_{12} \right) + \sin^{4}\theta_{13}. 
\end{eqnarray}
In Eq. (\ref{equation:oscillation2}) we have neglected matter effects, which modify the survival probability by $<$1\%~\cite{Giunti2007}. From a global analysis of neutrino oscillation data involving solar, accelerator, and reactor neutrinos, including the present KamLAND data, we obtain $P_{ee} = 0.551 \pm 0.015$.  The less-than 3\% uncertainty in $P_{ee} $ is negligible compared to the statistical uncertainty of KamLAND's current geo $\overline{\nu}_{e}$ flux measurement.

\begin{figure}[t]
\begin{center}
\includegraphics[angle=0,width=1.0\columnwidth]{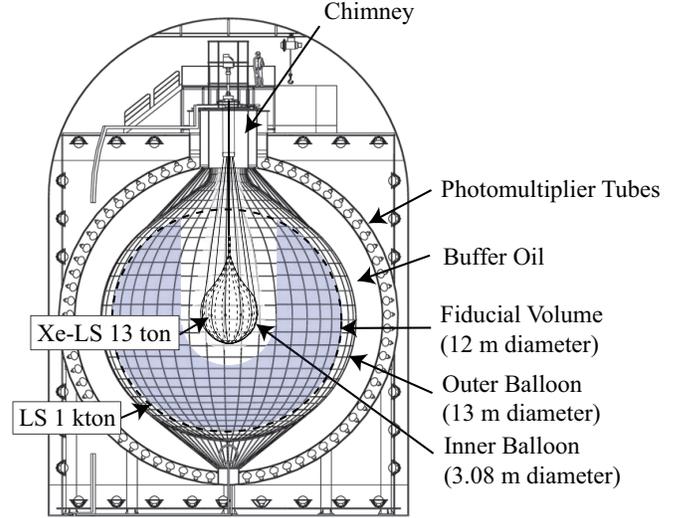}
\vspace{-0.8cm}
\end{center}
\caption[]{Schematic diagram of the KamLAND detector. The shaded region in the liquid scintillator indicates the volume for the $\overline{\nu}_{e}$ analysis after the inner balloon was installed.}
\label{figure:detector}
\end{figure}

\section{The KamLAND Experiment}
\label{section:Experiment}

KamLAND is located in Gifu Prefecture, Japan, under Mount Ikenoyama at a depth of $\sim$2700\,m water-equivalent. The primary volume consists of 1~kton of ultra-pure liquid scintillator~(LS), which comprises the neutrino interaction target (Fig.~\ref{figure:detector}). The LS is contained in  a 13-m-diameter spherical balloon made of 135-$\mu$m-thick transparent nylon/EVOH (ethylene vinyl alcohol copolymer) composite film. The balloon is suspended in non-scintillating purified mineral oil contained inside an 18-m-diameter stainless steel tank. The LS consists of 80\% dodecane and 20\% pseudocumene (1,2,4-trimethylbenzene) by volume, and $1.36 \pm 0.03$~g/liter PPO (2,5-diphenyloxazole) as a fluor. The scintillation light is viewed by an array of 1325 specially-developed fast 20-inch-diameter photomultiplier tubes (PMTs) masked to 17-inch diameter, and 554 older unmasked 20-inch PMTs, providing 34\% solid-angle coverage in total. This inner detector (ID) is surrounded by a \mbox{3.2~kton} water-Cherenkov outer detector~(OD) that serves as a cosmic-ray muon veto counter. 

In September 2011, the \mbox{KamLAND-Zen} neutrinoless double beta-decay search was launched~\cite{Gando2012}.  This search makes use of KamLAND's extremely low background and  suspends a $\beta\beta$ source, 13 tons of Xe-loaded liquid scintillator (Xe-LS), in a 3.08-m-diameter inner balloon (IB) at the center of the detector, as shown in Fig.~\ref{figure:detector}. To avoid backgrounds from the IB and its support material, the $\overline{\nu}_{e}$ analysis reported here is restricted to events occurring well outside the IB.

Electron antineutrinos are detected through the inverse $\beta$-decay reaction, \mbox{$\overline{\nu}_{e}+p\rightarrow e^{+}+n$}, which yields a delayed coincidence~(DC) event pair signature that provides a powerful tool to suppress backgrounds. The prompt scintillation light from the $e^{+}$ gives a measure of the incident $\overline{\nu}_{e}$ energy, \mbox{$E_{\nu} \simeq E_{\rm p} + \overline{E}_{n} + 0.8 ~{\rm MeV}$, where $E_{\rm p}$} is the sum of the $e^{+}$ kinetic energy and annihilation $\gamma$ energies, and $\overline{E}_{n}$ is the average neutron recoil energy, $O(10 ~\rm{keV})$. The mean time for capture of the neutron in the LS is $207.5 \pm 2.8 ~\mu \rm{s}$~\cite{Abe2010}. The scintillation light from the capture $\gamma$ constitutes the delayed event of the DC pair.

\begin{center}
\begin{table*}[t]
\caption{\label{table:background}Estimated backgrounds for $\overline{\nu}_{e}$ in the energy range between $0.9\,{\rm MeV}$ and $8.5\,{\rm MeV}$ after event selection cuts.
}
\centering
\begin{tabular}{@{}*{22}{llr@{}l@{}c@{}r@{}lr@{}l@{}c@{}r@{}lr@{}l@{}c@{}r@{}lr@{}l@{}c@{}r@{}l}}
\hline
\hline
\multicolumn{2}{l}{Background}  & \multicolumn{5}{c}{~~Period 1~~~~~~~~~~~} & \multicolumn{5}{c}{Period 2~~~~~~~~~~~} & \multicolumn{5}{c}{Period 3~~~~~~~~~~~} & \multicolumn{5}{c}{~~All Periods~~~~~} \\
 \multicolumn{2}{l}{} & \multicolumn{5}{c}{~~(1486 days)~~~~~~~~~~~} & \multicolumn{5}{c}{(1154 days)~~~~~~~~~~~} & \multicolumn{5}{c}{(351 days)~~~~~~~~~~~} & \multicolumn{5}{c}{~~(2991 days)~~~~~} \\
\hline
1 & Accidental & 76.&1 & ~$\pm$~ & 0.&1 & 44.&7 & ~$\pm$~ & 0.&1 & 4.&7 & ~$\pm$~ & 0.&1 & 125.&5 & ~$\pm$~ & 0.&1\\
2 & $^{9}$Li/$^{8}$He & 17.&9 & $\pm$ & 1.&4 & 11.&2 & $\pm$ & 1.&1 & 2.&5 & $\pm$ & 0.&5 & 31.&6 & $\pm$ & 1.&9\\
\multirow{2}{*}{3 $\bigg{\lbrace}$}  
& $^{13}{\rm C}(\alpha,{\it n})^{16}{\rm O}_{\rm g.s.}$, elastic scattering & 160.&4 & ~$\pm$~ & 16.&4 & 16.&5 & ~$\pm$~ & 3.&8 & 2.&3 & ~$\pm$~ & 1.&0 & 179.&0 & ~$\pm$~ & 21.&1\\
& $^{13}{\rm C}(\alpha,{\it n})^{16}{\rm O}_{\rm g.s.}$, $^{12}{\rm C}({\it n},{\it n'})^{12}{\rm C}^{*}$ (4.4 MeV $\gamma$) & 6.&9 & ~$\pm$~ & 0.&7 & 0.&7 & ~$\pm$~ & 0.&2 & 0.&10 & ~$\pm$~ & 0.&04 & 7.&7 & ~$\pm$~ & 0.&9\\
\multirow{2}{*}{4 $\bigg{\lbrace}$} 
& $^{13}{\rm C}(\alpha,{\it n})^{16}{\rm O^{*}}$, 1st e.s. (6.05 MeV $e^{+}e^{-}$) & 14.&6 & ~$\pm$~ & 2.&9 & 1.&7 & ~$\pm$~ & 0.&5 & 0.&21 & ~$\pm$~ & 0.&09 & 16.&5 & ~$\pm$~ & 3.&5\\
& $^{13}{\rm C}(\alpha,{\it n})^{16}{\rm O^{*}}$, 2nd e.s. (6.13 MeV $\gamma$) & 3.&4 & ~$\pm$~ & 0.&7 & 0.&4 & ~$\pm$~ & 0.&1 & 0.&05 & ~$\pm$~ & 0.&02 & 3.&9 & ~$\pm$~ & 0.&8\\
5 & Fast neutron and atmospheric neutrino~~~~~ & \multicolumn{5}{c}{$<$ 7.7}~~~~~~~ & \multicolumn{5}{c}{$<$ 5.9}~~~~~~~ & \multicolumn{5}{c}{$<$ 1.7}~~~~~~~~~ & \multicolumn{5}{c}{$<$ 15.3}~~\\
\hline
Total  \hspace{-0.3cm} &  & 279.&2 & $\pm$ & 22.&1 & 75.&2 & $\pm$ & 7.&6 & 9.&9 & $\pm$ & 2.&1 & 364.&1 & $\pm$ & 30.&5\\
\hline
\hline
\end{tabular}
\end{table*}
\end{center}

\begin{figure*}
\begin{center}
\includegraphics[angle=270,width=165mm]{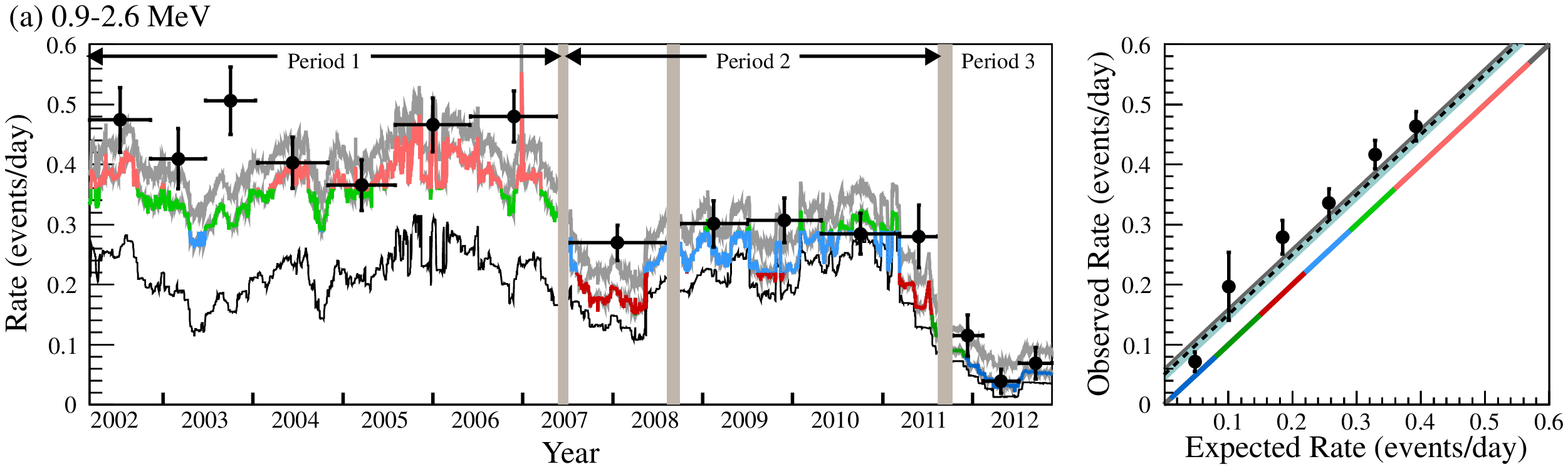}
\includegraphics[angle=270,width=165mm]{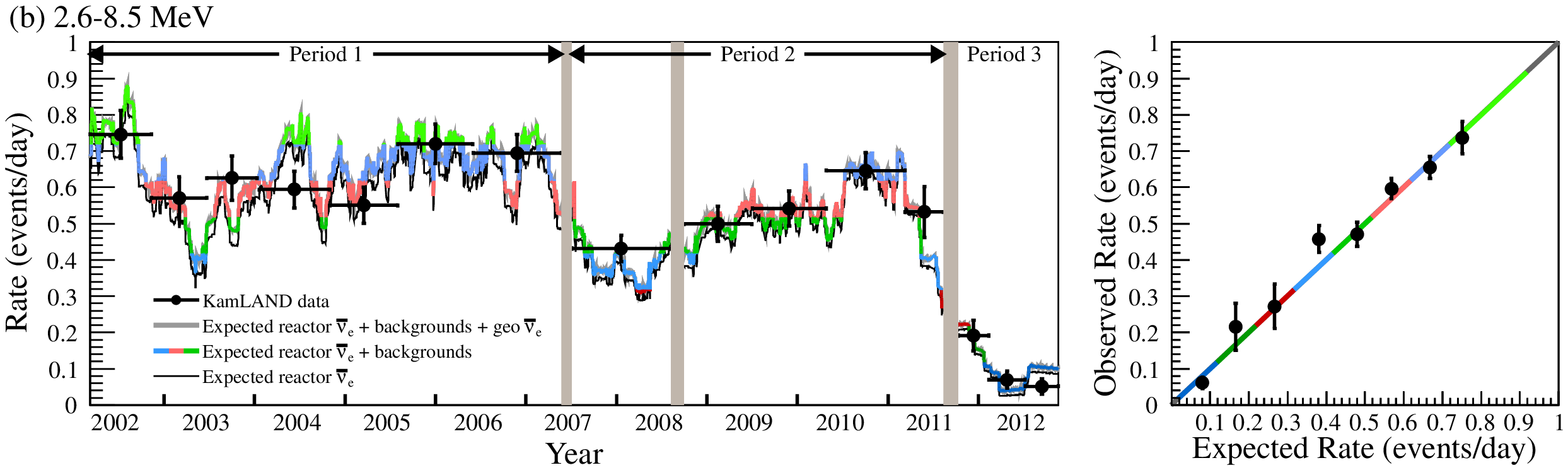}
\end{center}
\caption[]{Time evolution of expected and observed rates at KamLAND for $\overline{\nu}_{e}$'s with energies between (a) $0.9\,{\rm MeV}$ and $2.6\,{\rm MeV}$ and (b) $2.6\,{\rm MeV}$ and $8.5\,{\rm MeV}$. The points indicate the measured rates in a coarse time binning, while the curves show the expected rate variation for reactor $\overline{\nu}_{e}$'s (black line), reactor $\overline{\nu}_{e}$'s $+$ backgrounds (colored line), and reactor $\overline{\nu}_{e}$'s $+$ backgrounds $+$ geo $\overline{\nu}_{e}$'s (gray line). The geo $\overline{\nu}_{e}$ rates are calculated from the reference model~\cite{Enomoto2007}. The vertical bands correspond to data periods not used in the analysis. In the right panel of (a), the data are grouped according to periods of similar expected reactor $\overline{\nu}_{e}$ + background rates, as denoted by the colored bands. The observed event rate for each group is plotted at the exposure-weighted expected event rate within the group. 
The efficiency-corrected best-fit value of the geo $\overline{\nu}_{e}$ rate from the full spectral analysis (dashed line), its 1$\sigma$ error (shaded region), and the model expectation (gray line) are drawn for comparison. The contribution of geo $\overline{\nu}_{e}$'s in (b) is negligible. The oscillation parameters used to calculate the expected reactor $\overline{\nu}_{e}$ rate are the best-fit values from the global oscillation analysis: $\tan^{2} \theta_{12} = 0.436^{+0.029}_{-0.025}$, $\Delta m^{2}_{21} = 7.53^{+0.18}_{-0.18} \times 10^{-5}\,{\rm eV}^{2}$, and \mbox{$\sin^{2} \theta_{13} = 0.023^{+0.002}_{-0.002}$}.}
\label{figure:time}
\end{figure*}

\vspace{-1.0cm}

\section{Antineutrino candidate event selection}
\label{section:Experiment}

The data reported here are based on a total live-time of 2991~days, collected between March 9, 2002 and November 20, 2012. The data set is divided into three periods. \mbox{Period 1} (1486~days live-time) refers to data taken up to May 2007, at which time we embarked on a LS purification campaign that continued into 2009.  \mbox{Period 2} (1154~days live-time) refers to data taken during and after the LS purification campaign, and \mbox{Period 3} (351~days live-time) denotes the data taken after installing the IB. We removed periods of low data quality and high dead time that occurred during LS purification and KamLAND-Zen IB installation. The LS purification reduced the dominant \mbox{Period 1} background for $\bar{\nu}_{e}$'s, $^{13}{\rm C}(\alpha,{\it\ n})^{16}{\rm O}$ decays, by a factor of $\sim$20. The high-quality data taken after LS purification accounts for 50\% of the total live-time. Using a spherical fiducial scintillator volume with 6.0\,m radius, the number of target protons is estimated to be $(5.98 \pm 0.13) \times 10^{31}$, resulting in a total exposure of $(4.90 \pm 0.10) \times 10^{32}$ target-proton-years. The reduced fiducial volume in \mbox{Period 3} is accounted for in the detection efficiency; it contributes negligible additional fiducial volume uncertainty for \mbox{Period 3}.

Event vertex and energy reconstruction is based on the timing and charge distributions of scintillation photons recorded by the ID PMTs. The reconstruction is calibrated with $^{60}$Co, $^{68}$Ge, $^{203}$Hg, $^{65}$Zn, $^{241}$Am$^{9}$Be, $^{137}$Cs, and $^{210}$Po$^{13}$C radioactive sources. The achieved vertex resolution is \mbox{$\sim$12\,cm / $\sqrt{E(\rm{MeV})}$}, and the energy resolution is \mbox{6.4\% / $\sqrt{E(\rm{MeV})}$}. The nonlinear, particle-dependent conversion between deposited (real) energy and KamLAND's prompt energy scale is performed with a model incorporating Birks quenching and Cherenkov emission. The model parameters are constrained with calibration data, and contribute a 1.8\% systematic uncertainty to the measured value of $\Delta m^2_{21}$. Using calibration data taken throughout the fiducial volume during Period 1, we find that the deviation of reconstructed vertices from the actual deployment locations is less than 3\,cm. Incorporating a study of muon-induced $^{12}$B/$^{12}$N decays~\cite{Berger2009}, the fiducial volume uncertainties are 1.8\% for the pre-purification data and 2.5\% for the post-purification data.

For the DC event pair selection, we apply the following series of cuts:  (i) prompt energy: \mbox{$0.9 < E_{\rm p} ({\rm MeV}) < 8.5;$} (ii) delayed energy: \mbox{$1.8 < E_{\rm d}({\rm MeV}) < 2.6$} (capture on $p$), or \mbox{$4.4 < E_{\rm d}({\rm MeV}) < 5.6$} (capture on $^{12}{\rm C}$); (iii) spatial correlation of prompt and delayed events: \mbox{$\Delta R < 2.0\,\rm{m};$} (iv) time separation between prompt and delayed events: \mbox{$0.5 < \Delta T (\mu{\rm s})<  1000;$} (v) fiducial volume radii: \mbox{$R_{\rm p}, R_{\rm d} < 6.0\,\rm{m};$} (vi) and for \mbox{Period 3}, delayed vertex position: \mbox{$R_{\rm d} > 2.5\,\rm{m}$} and \mbox{$\rho_{\rm d} > 2.5\,\rm{m}$}, \mbox{$Z_{\rm d} > 0\,\rm{m}$} (vertical central cylinder cut at the upper hemisphere) to eliminate backgrounds from the \mbox{KamLAND-Zen} material.
To maximize the sensitivity to $\overline{\nu}_{e}$ signals, we perform an additional event selection designed to suppress accidental coincidence backgrounds from radioactive contaminants in the detector while maintaining high efficiency for $\overline{\nu}_{e}$'s. First, using a combination of Monte Carlo, data-driven, and analytical methods, we constructed a probability density function (PDF) for the $\overline{\nu}_{e}$ signal ($f_{\overline{\nu}_{e}}$) and accidental ($f_{acc}$) coincidence events. The PDF is based on the six cut parameters ($E_{\rm p}$, $E_{\rm d}$, $\Delta R$, $\Delta T$, $R_{\rm p}$, $R_{\rm d}$). For each candidate pair, we calculate the discriminant $L = \frac{f_{\overline{\nu}_{e}}}{f_{\overline{\nu}_{e}} + f_{acc}}$ and determine a selection value, $L_{cut}(E_{\rm p})$, to maximize the figure-of-merit $\frac{S}{\sqrt{S + B_{acc}}}$ for prompt energy intervals of 0.1\,MeV. In the figure-of-merit, $S$ is the number of the expected signal events assuming an oscillation-free reactor spectrum and the geo $\overline{\nu}_{e}$ fluxes predicted by~\cite{Enomoto2007}. $B_{acc}$ corresponds to the number of accidental background events, as measured using an out-of-time delayed coincidence window selection ($10\,{\rm ms} < \Delta T < 20\,{\rm s}$). The selection efficiency is calculated via Monte Carlo from the ratio of selected $\overline{\nu}_{e}$'s to the total number of generated $\overline{\nu}_{e}$'s in $R < 6\,{\rm m}$. The systematic uncertainty is evaluated using $^{68}$Ge and $^{241}$Am$^{9}$Be source calibrations as discussed in~\cite{Gando2011a}. The total number of events passing all selection criteria is 2611.

The reactor fluxes can be predicted from reactor operation records, which are provided to the KamLAND Collaboration by a consortium of Japanese electric power companies, and include the thermal power variation as well as fuel exchange and reshuffling data for all Japanese commercial reactors. The thermal power generation used for the normalization of the fission rates is measured to within 2\%. Only four isotopes contribute significantly to the $\overline{\nu}_{e}$ emission spectra; the relative fission yields, averaged over the entire live-time period for this result, are (0.567 : 0.078 : 0.298 : 0.057) for ($^{235}$U : $^{238}$U : $^{239}$Pu : $^{241}$Pu), respectively. A recent recalculation of the $\overline{\nu}_{e}$ spectra per fission of these isotopes introduced a $\sim$3\% upward shift~\cite{Mueller2011,Huber2011} relative to the previous standard calculation~\cite{Schreckenbach1985,Hahn1989}, causing past measurements at short-baselines to appear to have seen fewer $\bar{\nu}_{e}$'s  than expected. It has been speculated that this so-called Reactor Antineutrino Anomaly may be due to some systematic uncertainty or bias, or could potentially be due to oscillation into a heavy sterile neutrino state with $\Delta m^{2} \sim 1\,{\rm eV}^{2}$~\cite{Mention2011}.  To make our analysis insensitive to these effects, the normalization of the cross section per fission for each reactor is adjusted to reproduce the Bugey4 result~\cite{Declais1994}:
\begin{eqnarray}
\label{equation:scaling}
\left< \sigma \right>_{reac.} = \left< \sigma \right>_{Bugey4} + \sum_{i} (\alpha_{i}^{reac.} - \alpha_{i}^{Bugey4}) \left< \sigma \right>_{i}
\end{eqnarray}
where $\alpha_{i}$ is the fractional fission rate of the isotope $i$. The contribution from Korean reactors, based on reported electric power generation, is estimated to be $(4.9 \pm 0.5)\%$. The contribution from Japanese research reactors and all other reactors around the world is $(1.1 \pm 0.6)\%$. The levels of the long-lived, out-of-equilibrium fission products $^{90}$Sr, $^{106}$Ru, and $^{144}$Ce~\cite{Kopeikin2001} are evaluated from the history of fission rates for each isotope and are found to contribute an additional $(0.7 \pm 0.3)\%$. Applying the selection cut efficiency, we expect a total of $3564 \pm 145$ events from reactors in the absence of $\overline{\nu}_{e}$ disappearance.

A calculation of the geo $\overline{\nu}_{e}$ flux at KamLAND based on the reference Earth model of~\cite{Enomoto2007} gives an expected 109 and 27 geo $\overline{\nu}_{e}$ events from U and Th, respectively. Since the estimation of the geo $\overline{\nu}_{e}$ yield is highly model-dependent, the event rates from the U and Th decay chains are not constrained in the oscillation analysis. Only the prompt energy spectral shapes, which are independent of the Earth model, are used to constrain their contributions. A possible contribution from a hypothetical reactor-$\overline{\nu}_{e}$ source at the Earth's center, motivated by \cite{Herndon2003} and investigated in \cite{Bellini2010} and \cite{Gando2011b}, is neglected as a background in the fit for the oscillation parameters and geoneutrino fluxes, but is addressed briefly below as an independent signal.

In \mbox{Period 1}, the dominant background is the $^{13}{\rm C}(\alpha,{\it\ n})^{16}{\rm O}$ reaction, generated from the $\alpha$-decay of $^{210}$Po in the LS. 
The neutrons in this reaction are produced with energies up to 7.3\,MeV, but the visible energy is quenched to below 2.7\,MeV. Accounting for the energy-dependent efficiency of the $L_{cut}(E_{\rm p})$ selection, the estimated number of $^{13}{\rm C}(\alpha,{\it\ n})^{16}{\rm O}$ background events is $207.1 \pm 26.3$  in the energy region $0.9 < E_{\rm p} ({\rm MeV}) < 8.5$. The accidental background, which dominates in \mbox{Periods 2} and 3, is measured with an out-of-time delayed coincidence window from 10\,ms to 20\,s to be $125.5 \pm 0.1$ events. Including smaller contributions from cosmogenically produced radioactive isotopes, fast neutrons produced by cosmic-ray muons, and atmospheric neutrinos, the total background is estimated to be $364.1 \pm 30.5$ events. The backgrounds are detailed in Table~\ref{table:background}.

\begin{figure}[t!]
\begin{center}
\hspace{-0.4cm}
\includegraphics[angle=0,width=80mm]{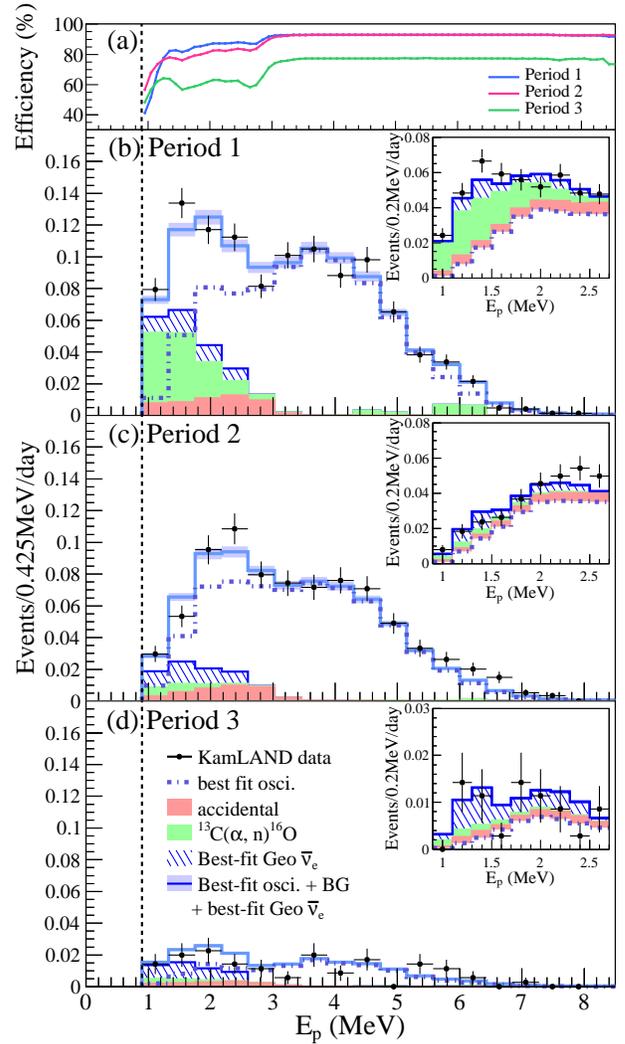}
\hspace{-0.8cm}
\end{center}
\caption[]{Prompt energy spectrum of $\overline{\nu}_{e}$ candidate events above the 0.9\,MeV energy threshold (vertical dashed line) for each data taking period. The background, reactor and geo $\overline{\nu}_{e}$ contributions are the best-fit values from a KamLAND-only analysis. The prompt energy spectra of $\overline{\nu}_{e}$ candidate events in the low-energy region are also shown in the inset panels with a finer binning. The top panel shows the energy-dependent selection efficiency curves for each period.}
\label{figure:energy_spectrum_period_0.9_8.5MeV}
\end{figure}

\begin{figure}[t]
\vspace{0.3cm}
\begin{center}
\hspace{-0.3cm}
\includegraphics[angle=270,width=0.90\columnwidth]{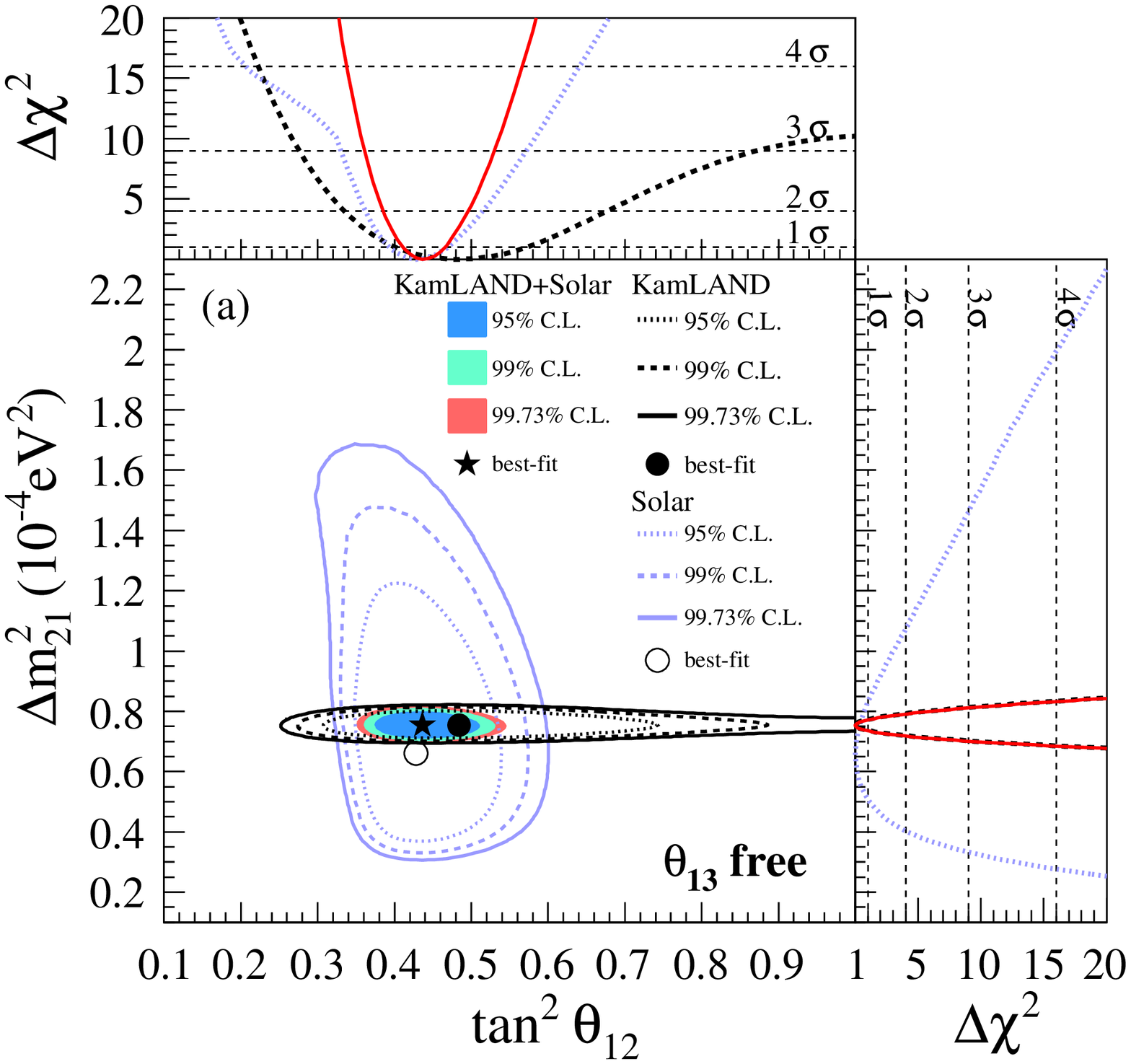}
\vspace{0.2cm}
\end{center}
\begin{center}
\includegraphics[angle=270,width=0.91\columnwidth]{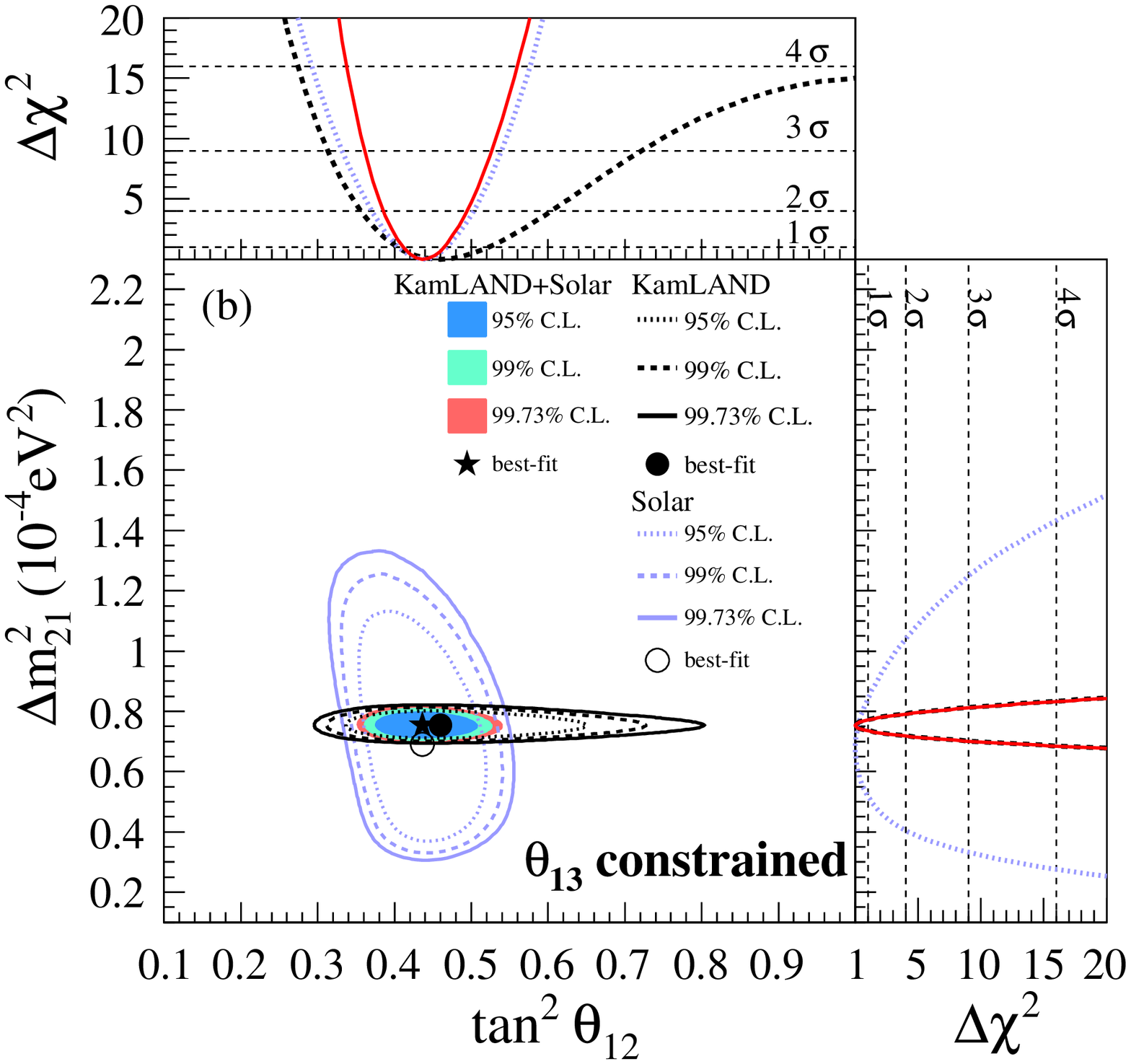}
\vspace{-0.2cm}
\end{center}
\caption[]{Allowed regions projected in the ($\tan^{2} \theta_{12}$, $\Delta m^{2}_{21}$) plane, for solar and KamLAND data from the three-flavor oscillation analysis for (a) $\theta_{13}$ free and (b) $\theta_{13}$ constrained by accelerator and short-baseline reactor neutrino experiments. The shaded regions are from the combined analysis of the solar and KamLAND data. The side panels show the $\Delta \chi^{2}$-profiles projected onto the $\tan^{2} \theta_{12}$ and $\Delta m^{2}_{21}$ axes.}
\label{figure:oscillation_contour}
\end{figure}

\begin{figure}[h]
\vspace{0.3cm}
\begin{center}
\hspace{-0.3cm}
\includegraphics[angle=270,width=0.90\columnwidth]{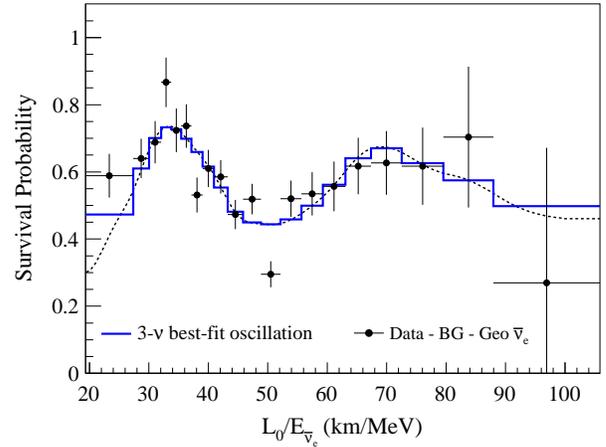}
\vspace{-0.2cm}
\end{center}
\caption[]{Ratio of the observed $\overline{\nu}_{e}$ spectrum to the expectation for no-oscillation versus $L_{0}/E$ for the KamLAND data. $L_{0} = 180\,{\rm km}$ is the flux-weighted average reactor baseline.  The 3-$\nu$ histogram is the best-fit survival probability curve from the three-flavor unbinned maximum-likelihood analysis using only the KamLAND data.}
\vspace{-5mm}
\label{figure:LE}
\end{figure}

\begin{figure}[t]
\begin{center}
\hspace{-0.2cm}
\includegraphics[angle=0,width=0.95\columnwidth]{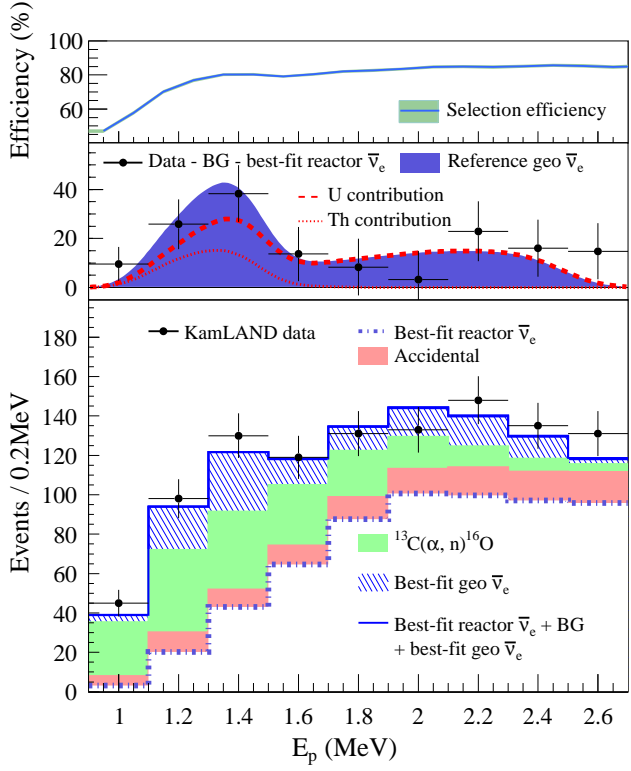}
\vspace{-0.4cm}
\end{center}
\caption[]{Prompt energy spectrum of the $\overline{\nu}_{e}$ events in the low-energy region for all data taking periods. Bottom panel: data together with the best-fit background and geo $\overline{\nu}_{e}$ contributions. The fit incorporates all available constraints on the oscillation parameters. The shaded background and geo $\overline{\nu}_{e}$ histograms are cumulative. Middle panel: observed geo $\overline{\nu}_{e}$ spectrum after subtraction of reactor $\overline{\nu}_{e}$'s and other background sources. The dashed and dotted lines show the best-fit U and Th spectral contributions, respectively. The blue shaded curve shows the expectation from the geological reference model of~\cite{Enomoto2007}. Top panel: the energy-dependent selection efficiency.}
\label{figure:spectrum}
\end{figure}

\begin{figure}[t]
\vspace{0.3cm}
\begin{center}
\hspace{-0.3cm}
\includegraphics[angle=270,width=0.90\columnwidth]{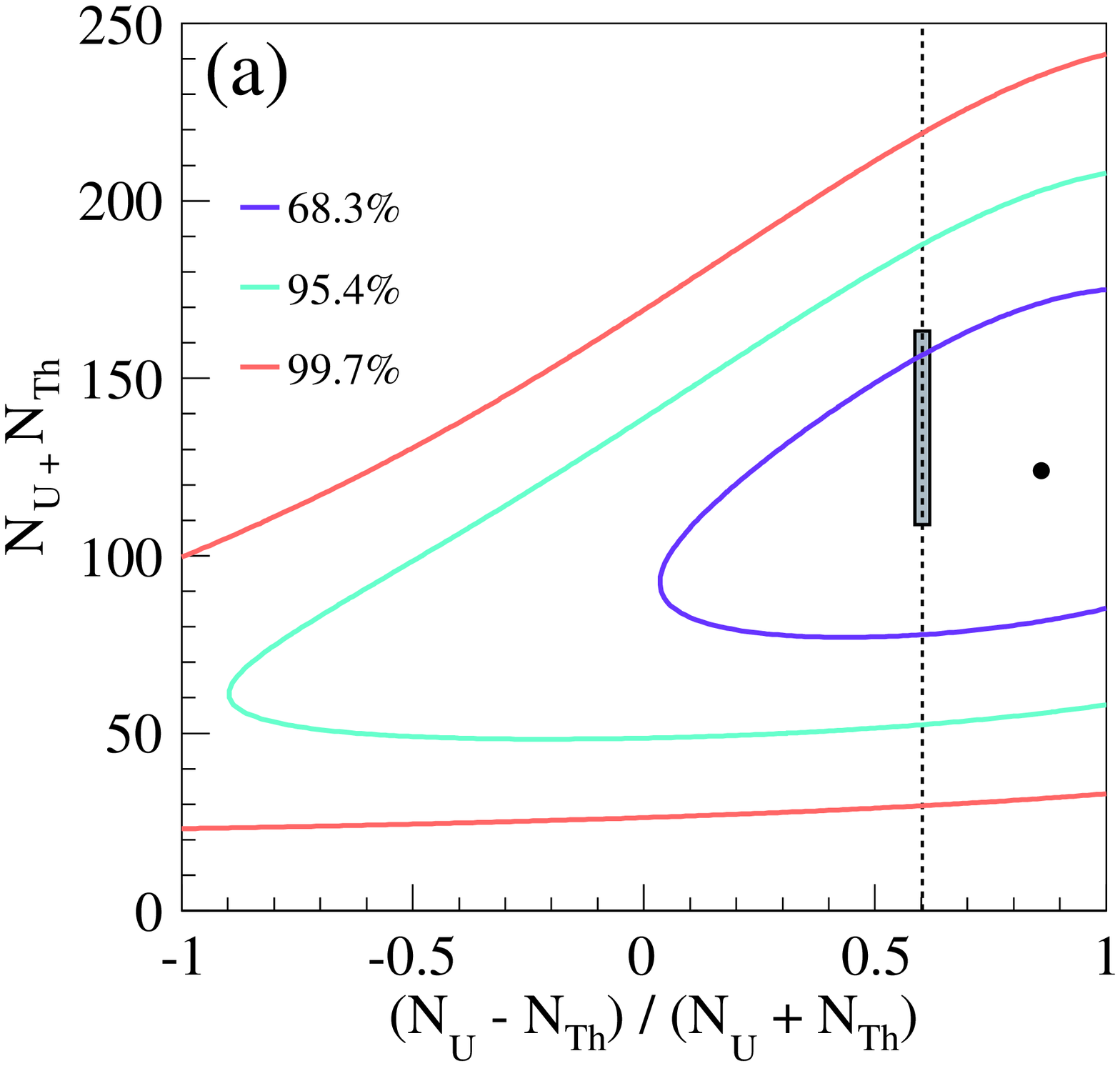}
\vspace{0.2cm}
\end{center}
\begin{center}
\includegraphics[angle=270,width=0.91\columnwidth]{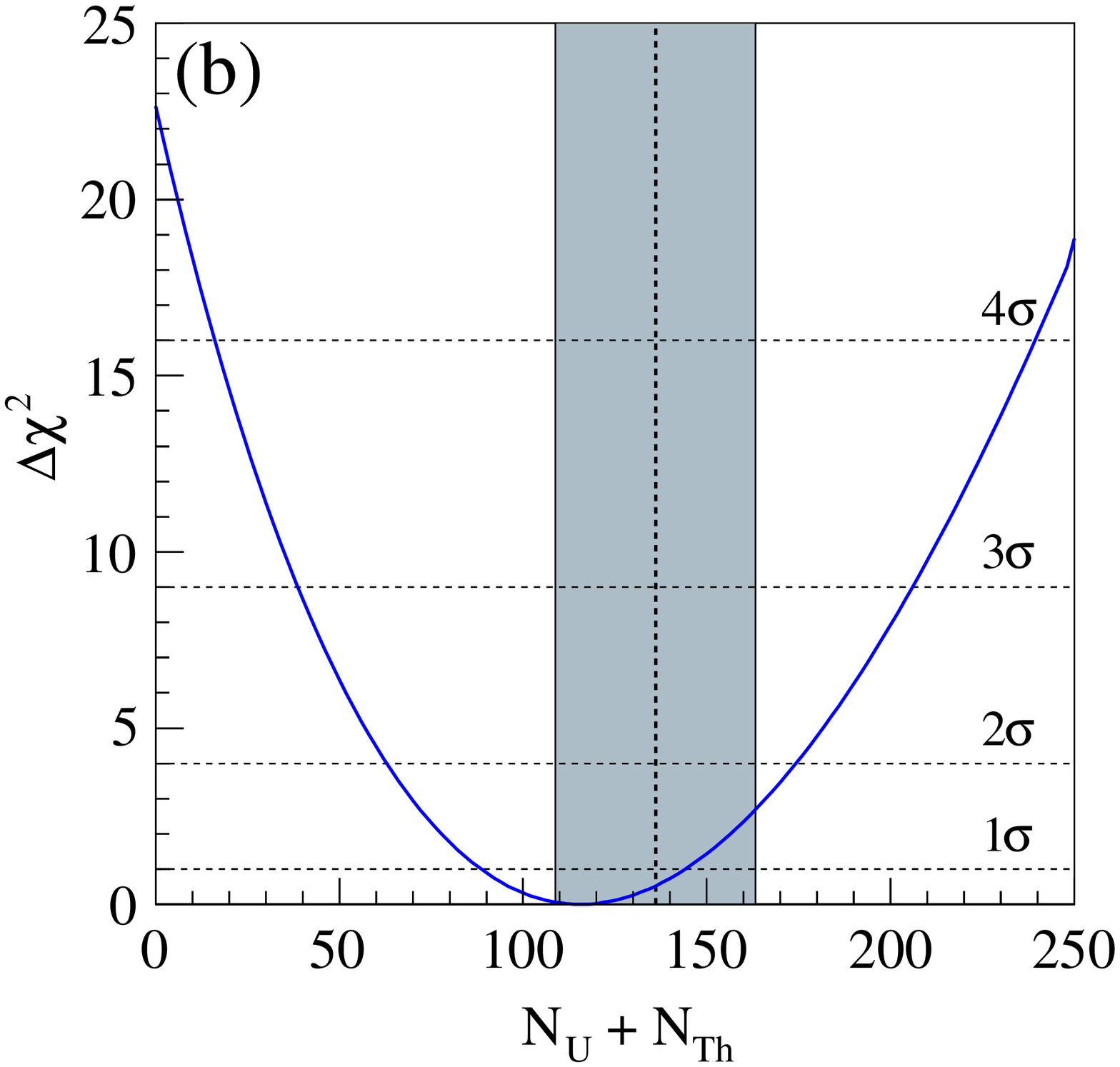}
\vspace{-0.2cm}
\end{center}
\caption[]{(a) Confidence level (C.L.) contours for the observed number of geo $\overline{\nu}_{e}$ events. The small shaded region represents the prediction of the reference model of~\cite{Enomoto2007}. The vertical dashed line represents the value of $(N_{\rm U} - N_{\rm Th}) / (N_{\rm U} + N_{\rm Th})$ expected for a Th/U mass ratio of 3.9 derived from chondritic meteorites. (b) $\Delta \chi^{2}$-profile from the fit to the total number of geo $\overline{\nu}_{e}$ events, fixing the Th/U mass ratio at 3.9. The grey band represent the geochemical model prediction, assuming a 20\% uncertainty in the abundance estimates.}
\vspace{-5mm}
\label{figure:confidence}
\end{figure}

\vspace{-0.2cm}

\section{Antineutrino Measurement and Oscillation Analysis}
\label{section:Flux}

To extract the neutrino oscillation parameters and geoneutrino fluxes, $\overline{\nu}_{e}$ candidates are analyzed with an unbinned maximum-likelihood method incorporating the event rate and the prompt energy spectrum shape, including their time variation, in the range $0.9 < E_{\rm p} ({\rm MeV}) < 8.5$. The $\chi^{2}$ is defined by 
\begin{eqnarray}
\label{equation:chi2}
\chi^{2} & = & \chi^{2}_{\rm rate}(\theta_{12}, \theta_{13}, \Delta m^{2}_{21}, N_{\rm BG1\rightarrow5}, N^{\rm{geo}}_{{\rm U},{\rm Th}}, \alpha_{\rm 1\rightarrow4}) \nonumber \\
& & - 2 \ln L_{\rm shape} (\theta_{12}, \theta_{13}, \Delta m^{2}_{21}, N_{\rm BG1\rightarrow5}, N^{\rm{geo}}_{{\rm U},{\rm Th}}, \alpha_{\rm 1\rightarrow4}) \nonumber \\
& & + \chi^{2}_{\rm BG}(N_{\rm BG1\rightarrow5}) + \chi^{2}_{\rm syst}(\alpha_{\rm 1\rightarrow4})  \nonumber \\
& & + \chi^{2}_{\rm osci}(\theta_{12}, \theta_{13}, \Delta m^{2}_{21})\;.
\end{eqnarray}
The terms are, in order: the $\chi^{2}$ contribution for (i) the time-varying event rate, (ii) the time-varying prompt energy spectrum shape, (iii) a penalty term for backgrounds, (iv) a penalty term for systematic uncertainties, and (v) a penalty term for the oscillation parameters. $N^{\rm{geo}}_{{\rm U},{\rm Th}}$ are the flux normalization parameters for U and Th geo $\overline{\nu}_{e}$'s, and allow for an Earth-model-independent analysis. $N_{\rm BG1\rightarrow5}$ are the expected number of backgrounds, and are allowed to vary in the fit but are constrained with the penalty term (iii) using the estimates described in the preceding section and listed, with the corresponding index, in Table~\ref{table:background}. $\alpha_{\rm 1\rightarrow4}$ parametrize the uncertainties on the reactor $\overline{\nu}_{e}$ spectrum, the energy scale, the event rate, and the energy-dependent detection efficiency; these parameters are allowed to vary in the analysis but are constrained by term (iv). Table~\ref{table:systematic} summarizes the systematic uncertainties on $\Delta m^{2}_{21}$ and the expected event rate of reactor $\overline{\nu}_{e}$'s. The overall rate uncertainties for \mbox{Period 1} and for \mbox{Periods 2 and 3} are 3.5\% and 4.0\%, respectively. Systematic uncertainties are conservatively treated as being fully correlated across all data taking periods. The penalty term (v) optionally provides a constraint on the neutrino oscillation parameters from solar~\cite{Cleveland1998,Abdurashitov2009,Bellini2011,Hosaka2006,Aharmim2011}, accelerator (T2K~\cite{Abe2011}, MINOS~\cite{Adamson2011}), and short-baseline reactor neutrino experiments (Double Chooz~\cite{Abe2012}, Daya Bay~\cite{An2013}, RENO~\cite{Ahn2012}).
 
\begin{center}
\begin{table}[b]
\caption{\label{table:systematic}Contributions to the systematic uncertainty in the neutrino oscillation parameters $\Delta m^{2}_{21}$, $\theta_{12}$, and $\theta_{13}$ for the earlier / later periods of measurement, denoted in the text as \mbox{Period 1 / Period 2 \& 3}. The overall uncertainties are 3.5\% / 4.0\% for \mbox{Period 1 / Period 2 \& 3}.
}
\begin{tabular}{@{}*{7}{l}}
\hline
\hline
 & Detector-related (\%) \hspace{-0.5cm} &  & Reactor-related (\%) \hspace{-0.5cm} & \\
\hline
$\Delta m^{2}_{21}$ \hspace{0.0cm} & Energy scale &  1.8 / 1.8  &  $\overline{\nu}_{e}$-spectra~\cite{Achkar1996} & \hspace{0.2cm} 0.6 / 0.6 \vspace{0.1cm}\\
Rate \hspace{0.2cm} & Fiducial volume & 1.8 / 2.5 &  $\overline{\nu}_{e}$-spectra~\cite{Declais1994} & \hspace{0.2cm} 1.4 / 1.4\\
 & Energy scale & 1.1 / 1.3 & Reactor power & \hspace{0.2cm} 2.1 / 2.1 \\
 & $L_{cut}(E_{\rm p})$ eff. & 0.7 / 0.8 & Fuel composition & \hspace{0.2cm} 1.0 / 1.0\\
 & Cross section & 0.2 / 0.2 & Long-lived nuclei & \hspace{0.2cm} 0.3 / 0.4\\
\hline
& Total \hspace{0.0cm} & 2.3 / 3.0 \hspace{0.2cm} & Total &\hspace{0.2cm} 2.7 / 2.8\\
\hline
\hline
\end{tabular}
\end{table}
\end{center}

\begin{center}
\begin{table}[b]
\caption{\label{table:fit-values}
Summary of the fit values for $\Delta m^{2}_{21}$, $\tan^{2} \theta_{12}$ and $\sin^{2} \theta_{13}$ from three-flavor neutrino oscillation analyses with various combinations of experimental data. }
\centering
\begin{tabular}{@{}*{4}{lccc}}
\hline
\hline
\hspace{0.3cm} Data combination & $\Delta m^{2}_{21}$ & \hspace{0.3cm} $\tan^{2} \theta_{12}$ \hspace{0.3cm} & \hspace{0.3cm} $\sin^{2} \theta_{13}$ \hspace{0.3cm} \\
\hline
KamLAND                                      & $7.54^{+0.19}_{-0.18}$ & $0.481^{+0.092}_{-0.080}$ & $0.010^{+0.033}_{-0.034}$ \\
KamLAND + solar                           & $7.53^{+0.19}_{-0.18}$ & $0.437^{+0.029}_{-0.026}$ & $0.023^{+0.015}_{-0.015}$ \\
KamLAND + solar + $\theta_{13}$ & $7.53^{+0.18}_{-0.18}$  & $0.436^{+0.029}_{-0.025}$ & $0.023^{+0.002}_{-0.002}$ \\
\hline
\hline
\end{tabular}
\end{table}
\vspace{-1.0cm}
\end{center}

Figure~\ref{figure:time} plots the time variation for the rates of reactor $\overline{\nu}_{e}$'s, geo $\overline{\nu}_{e}$'s, and backgrounds for the three data taking periods, assuming the best-fit oscillation parameters, and geo $\overline{\nu}_{e}$ fluxes from the reference model of~\cite{Enomoto2007}. Also drawn are the correlations between the measured and expected best-fit event rates, which should fit to a line with unit slope and zero offset in the absence of geo $\overline{\nu}_{e}$'s. The vertical displacement of the trend for events below 2.6 MeV is attributed to the contribution of geo $\overline{\nu}_{e}$'s.

Figure~\ref{figure:energy_spectrum_period_0.9_8.5MeV} shows the prompt energy spectra of $\overline{\nu}_{e}$ candidate events for each period. The reduction of the $^{13}{\rm C}(\alpha,{\it\ n})^{16}{\rm O}$ background in \mbox{Period 2} and of reactor $\overline{\nu}_{e}$'s in \mbox{Period 3} can clearly be seen. For the three-flavor KamLAND-only analysis ($\chi^{2}_{\rm osci} = 0$), the fit oscillation parameter values are  $\Delta m^{2}_{21} = 7.54^{+0.19}_{-0.18} \times 10^{-5}\,{\rm eV}^{2}$, $\tan^{2} \theta_{12} = 0.481^{+0.092}_{-0.080}$, and \mbox{$\sin^{2} \theta_{13} = 0.010^{+0.033}_{-0.034}$}. The contours are nearly symmetric about $\tan^{2} \theta_{12} = 1$, but the best-fit values for $\tan^{2} \theta_{12} > 1$ are slightly disfavored over those for $\tan^{2} \theta_{12} < 1$, with $\Delta \chi^{2}$ = 0.8. Assuming {\it CPT} invariance, the oscillation parameter values from a combined analysis including constraints from solar neutrino experiments are $\tan^{2} \theta_{12} = 0.437^{+0.029}_{-0.026}$, $\Delta m^{2}_{21} = 7.53^{+0.19}_{-0.18} \times 10^{-5}\,{\rm eV}^{2}$, and \mbox{$\sin^{2} \theta_{13} = 0.023^{+0.015}_{-0.015}$}. A global analysis including also constraints on $\theta_{13}$ from accelerator and short-baseline reactor neutrino experiments yields $\tan^{2} \theta_{12} = 0.436^{+0.029}_{-0.025}$, $\Delta m^{2}_{21} = 7.53^{+0.18}_{-0.18} \times 10^{-5}\,{\rm eV}^{2}$, and \mbox{$\sin^{2} \theta_{13} = 0.023^{+0.002}_{-0.002}$}. The fit values for the different combinations are summarized in Table~\ref{table:fit-values}. Figure~\ref{figure:oscillation_contour} shows the extracted confidence intervals in the ($\tan^{2}\theta_{12}$, $\Delta m^{2}_{21}$) plane with and without the $\theta_{13}$ constraint.

The KamLAND data illustrates the oscillatory shape of reactor $\overline{\nu}_{e}$'s arising from neutrino oscillation. The ratio of the background- and geo-$\overline{\nu}_{e}$-subtracted reactor $\overline{\nu}_{e}$ spectrum to the no-oscillation expectation is shown in Fig.~\ref{figure:LE} as a function of $L_{0}/E$, where $L_{0} = 180\,\rm{km}$ is the flux-weighted average reactor baseline. The improved determination of the geo $\overline{\nu}_{e}$ flux resulting from the addition of the reactor-off data makes the second peak at $L_{0}/E = 70\,{\rm km/MeV}$ more evident than in previous analyses.

For the geo $\overline{\nu}_{e}$ flux measurement we incorporate all available constraints on the oscillation parameters. The insets in Fig.~\ref{figure:energy_spectrum_period_0.9_8.5MeV} detail the observed spectra in the low-energy region for each data taking period. Figure~\ref{figure:spectrum} shows the measured geo $\overline{\nu}_{e}$ event spectrum after subtracting the best-fit reactor $\overline{\nu}_{e}$ and background spectra. The best-fit to the unbinned data yields 116 and 8 geo $\overline{\nu}_{e}$'s from U and Th decays, respectively. The joint confidence intervals for the sum $N_{\rm U} + N_{\rm Th}$ and the asymmetry factor $(N_{\rm U} - N_{\rm Th}) / (N_{\rm U} + N_{\rm Th})$ are shown in Fig.~\ref{figure:confidence}. This result agrees with the expectation from the geological reference model of~\cite{Enomoto2007}. We obtained an upper limit of $<$19 (90\% C.L.) in the ${\rm Th}/{\rm U}$ mass ratio, indicating the separation of U and Th $\overline{\nu}_{e}$'s. Assuming a Th/U mass ratio of 3.9, as predicted by the geochemical model of~\cite{McDonough1995} from the abundances observed in chondritic meteorites, the total number of U and Th geo $\overline{\nu}_{e}$ events is $116^{+28}_{-27}$, with a $\Delta \chi^{2}$-profile as shown in Fig.~\ref{figure:confidence}(b). This result corresponds to an (oscillated) $\overline{\nu}_{e}$ flux of \mbox{$3.4^{+0.8}_{-0.8} \times 10^{6}\,{\rm cm^{-2}s^{-1}}$} at KamLAND, or a total antineutrino flux including all flavors of \mbox{$6.2^{+1.5}_{-1.5} \times 10^{6}\,{\rm cm^{-2}s^{-1}}$}. From the $\Delta \chi^{2}$-profile (Fig.~\ref{figure:confidence}(b)), we find that the null hypothesis is disfavored with a $p$-value of $2 \times 10^{-6}$.

The KamLAND data also tests the hypothesis of a natural nuclear reactor in the Earth's core~\cite{Herndon2005} assuming a constant power output over the duration of the experiment. The oscillation parameters are constrained from the solar, accelerator, and reactor neutrino data, while the contributions from geological reactor $\overline{\nu}_{e}$'s and from U and Th geo $\overline{\nu}_{e}$'s are allowed to vary. The fit gives a limit on the geological reactor power of $<$3.1\,TW at 90\% C.L. ($<$3.7\,TW at 95\% C.L.), an improvement of a factor of 1.7 over the previous KamLAND result~\cite{Gando2011b}, due primarily to the reduction of the commercial reactor $\overline{\nu}_{e}$ background in \mbox{Period 3}.

\section{Constraints on Earth models}
\label{section:Earth}

While the mantle is the most massive layer of the Earth's interior, its chemical composition is still uncertain. A quantitative estimate of the heat production by radiogenic components is of particular importance for understanding dynamic processes such as mantle convection. Indeed, precisely how the mantle convects is still not fully understood, and controversy remains as to whether two-layer convection or whole-volume convection provides a more accurate description. In this work, we carry out a comparison of existing Earth models using the KamLAND geo $\overline{\nu}_{e}$ data on the basis of simple but appropriate assumptions. 

The crustal contribution to the flux at KamLAND can be estimated from compositional data through rock sampling~\cite{Enomoto2007}. Since current Earth models predict that the lithophiles U and Th are absent in the core, for a first approximation of the radiogenic heat, we attribute any excess above the crustal contribution to U and Th uniformly distributed throughout the mantle. Under these generic assumptions, the measured KamLAND geo $\overline{\nu}_{e}$ flux translates to a total radiogenic heat production of $11.2^{+7.9}_{-5.1}$\,TW from U and Th. This calculation accounts for crustal uncertainties of 17\% and 10\% for U and Th, respectively, including correlated errors as suggested in~\cite{Fogli2006}. To parameterize the planetary-scale energy balance, the fraction of the global heat production from radioactive decays, the so-called ``Urey ratio'', is introduced. Allowing for mantle heat contributions of 3.0\,TW from other isotope decays~\cite{Arevalo2009, Enomoto2006}, we find that the convective Urey ratio, the contribution to the Urey ratio from just the mantle, is between 0.09 and 0.42 at 68\% C.L. This range favors models that allow for a substantial but not dominant contribution from the Earth's primordial heat supply.

\begin{figure}[t]
\begin{center}
\vspace{-0.4cm}
\includegraphics[angle=0,width=1.05\columnwidth]{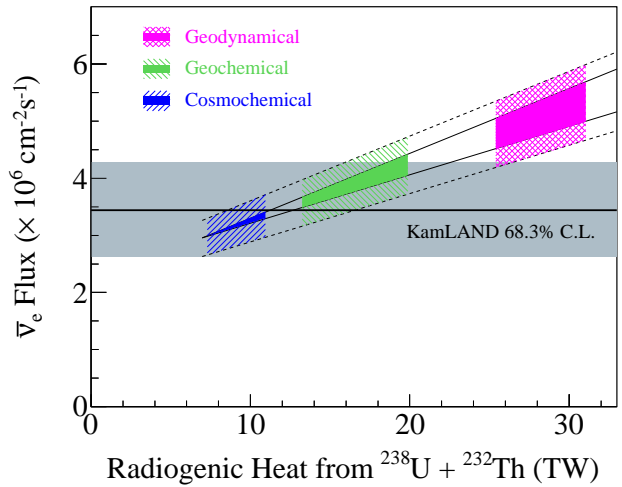}
\vspace{-0.5cm}
\end{center}
\caption[]{Geo $\overline{\nu}_{e}$ flux versus radiogenic heat from the decay chains of $^{238}$U and $^{232}$Th. The measured geo $\overline{\nu}_{e}$ flux (gray band) is compared with the expectations for the different mantle models from cosmochemical~\cite{Javoy2010}, geochemical~\cite{McDonough1995}, and geodynamical~\cite{Turcotte2002} estimates (color bands). The sloped band starting at 7\,TW indicates the response to the mantle $\overline{\nu}_{e}$ flux, which varies between the homogeneous and sunken-layer hypotheses (solid lines), discussed in the text. The upper and lower dashed lines incorporate the uncertainty in the crustal contribution.}
\label{figure:heat}
\end{figure}

Several established estimates of the BSE composition give different geo $\overline{\nu}_{e}$ flux predictions. Reference~\cite{Sramek2013} categorizes the models into three groups: geochemical, cosmochemical, and geodynamical. Geochemical models~\cite{McDonough1995}, such as the reference Earth model of~\cite{Enomoto2007}, use primordial compositions equal to those found in CI carbonaceous chondrites, but allow for elemental enrichment by differentiation, as deduced from terrestrial samples. Cosmochemical models~\cite{Javoy2010} assume a mantle composition similar to that of enstatite chondrites, and yield a lower radiogenic abundance. Geodynamical models~\cite{Turcotte2002}, on the other hand, require higher radiogenic abundances in order to drive realistic mantle convection. 

In Fig~\ref{figure:heat}, the observed geo $\overline{\nu}_{e}$ flux at KamLAND is compared with the expectations from these BSE compositional models assuming a common estimated crustal contribution~\cite{Enomoto2007}. The $\overline{\nu}_{e}$ flux predictions vary within the plotted vertical bands due to uncertainties in both the abundances of radioactive elements in the mantle as well as their distributions. The spread of the slope reflects the difference between two extreme radiochemical distributions: the ``homogeneous hypothesis'' in which U and Th are assumed to be distributed uniformly throughout the mantle, and the ``sunken-layer hypothesis'', which assumes that all of the U and Th below the crust collects at the mantle-core interface. While the statistical treatment of geological uncertainties is not straightforward, assuming Gaussian errors for the crustal contribution and for the BSE abundances, we find that the geodynamical prediction with the homogeneous hypothesis is disfavored at 89\% C.L. However, due to the limited statistical power of the data, all BSE composition models are still consistent within $\sim$2$\sigma$ C.L.

\section{Conclusion}
\label{section:Conclusion}

An updated KamLAND measurement of $\overline{\nu}_{e}$'s was presented. This data benefits from the significant reduction of reactor $\overline{\nu}_{e}$'s due to the long-term shutdown of commercial nuclear reactors in Japan. The geo $\overline{\nu}_{e}$ flux estimate is significantly improved by the reactor-off data.  Likewise, the reactor neutrino oscillation parameters are also better determined due to the reduction of uncertainties in the geo $\overline{\nu}_{e}$ flux and the rates of other backgrounds. Including constraints on $\theta_{13}$ from accelerator and short-baseline reactor neutrino experiments, a three-flavor analysis of solar and KamLAND data gives fit values for the oscillation parameters of $\tan^{2} \theta_{12} = 0.436^{+0.029}_{-0.025}$, $\Delta m^{2}_{21} = 7.53^{+0.18}_{-0.18} \times 10^{-5}\,{\rm eV}^{2}$, and \mbox{$\sin^{2} \theta_{13} = 0.023^{+0.002}_{-0.002}$}. Assuming a chondritic Th/U mass ratio of 3.9, we observed $116^{+28}_{-27}$ geo $\overline{\nu}_{e}$ events, which corresponds to a geo $\overline{\nu}_{e}$ flux of \mbox{$3.4^{+0.8}_{-0.8} \times 10^{6}\,{\rm cm^{-2}s^{-1}}$} at KamLAND. The observed rate is in agreement with the predictions from existing BSE composition models within $\sim$2$\sigma$ C.L. Currently, the ability of discriminating between models is limited by the experimental uncertainty. In the future, improved measurements with higher statistics and lower background can be achieved by larger detectors distant from commercial reactors. Multi-site flux data at a combination of crustal and oceanic  geological sites would be able to estimate the crustal contribution from a statistical correlation analysis and constrain mantle abundances more stringently.

\vspace{-0.5cm}

\section*{ACKNOWLEDGMENTS}

The \mbox{KamLAND} experiment is supported by the Grant-in-Aid for Specially Promoted Research under grant 21000001 of the Japanese Ministry of Education, Culture, Sports, Science and Technology; the World Premier International Research Center Initiative (WPI Initiative), MEXT, Japan; Stichting FOM in the Netherlands; and under the US Department of Energy (DOE) Grant No. DE-AC02-05CH11231, as well as other DOE grants to individual institutions. The Kamioka Mining and Smelting Company has provided service for activities in the mine.

\bibliography{AntiNeutrino}

\end{document}